\documentclass[journal, twocolumn]{IEEEtran}
\usepackage{cite}
\usepackage{amsmath,amssymb,amsfonts}
\usepackage{gensymb}

\usepackage{algorithmic}
\usepackage{graphicx}
\usepackage{textcomp}
\usepackage{xcolor}
\usepackage[hidelinks]{hyperref}
\usepackage{cancel}
\usepackage{array}
\usepackage{tablefootnote}
\usepackage{threeparttable}
\usepackage{epstopdf}

\def\BibTeX{{\rm B\kern-.05em{\sc i\kern-.025em b}\kern-.08em
    T\kern-.1667em\lower.7ex\hbox{E}\kern-.125emX}}

\DeclareMathOperator{\sinc}{sinc}
\DeclareMathOperator{\rect}{rect}

\begin{document}

\title{Aliased Time-Modulated Array OFDM Transmit System}

\author{
    Marcin Wachowiak,~\IEEEmembership{Student~Member,~IEEE,} 
    Kamil Yavuz Kapusuz,~\IEEEmembership{Senior~Member,~IEEE,}\\
    André Bourdoux,~\IEEEmembership{Senior~Member,~IEEE,}
    Sofie Pollin,~\IEEEmembership{Senior~Member,~IEEE,}%
    
    \thanks{
        Marcin Wachowiak and Sofie Pollin are with imec, 3001 Leuven, Belgium and also with the Katholieke Universiteit Leuven, 3000 Leuven, Belgium (e-mail: marcin.wachowiak@imec.be)}%
    \thanks{K.Y.~Kapusuz is with imec and the Department of Information Technology, IDLab/EM Group, Ghent University, Ghent, Belgium (e-mail:~kamilyavuz.kapusuz@ugent.be).}%
        
    \thanks{
        André Bourdoux is with imec, 3001 Leuven, Belgium. (Corresponding author: \textit{Marcin Wachowiak})}
}

\maketitle

\begin{abstract}
The time-modulated array is a simple array architecture in which each antenna is connected to an RF switch that serves as a modulator. The phase shift is achieved by digitally controlling the relative delay between the periodic modulating sequences of the antennas. Two factors limit the practical use of this architecture for communication and sensing. First, the switching frequency is high, as it must be a multiple of the sampling frequency. Second, the discrete modulating sequence introduces undesired harmonic replicas of the signal, which are out-of-band interference. This paper proposes the OFDM modulation with an appropriate precoder to facilitate the aliasing of the harmonic components to simultaneously reduce sideband radiation and switching frequency. The transmit signal has a repeated block structure in the frequency domain to facilitate coherent combining of the aliased signal blocks. As a result, a factor $A$ reduction in switching frequency is achieved at the cost of a factor $A$ reduction in communication capacity. Doubling $A$ reduces sideband radiation by around 2.9 dB. The feasibility of the proposed method is experimentally validated for wideband signals. Full-wave simulations are performed to validate the beamforming performance based on the experimental results.
\end{abstract}


\section{Introduction}

\subsection{Problem Statement}

Antenna arrays are an essential component of current and future generations of wireless systems~\cite{6g_vision}. However, when considering conventional architectures, the scaling of antenna arrays raises concerns about cost, complexity, and energy efficiency~\cite{scaling_digital_arrays, scaling_mmw_arrays}. To address scaling problems, alternative array architectures should be considered~\cite{unconvent_array_arch}.
A time-modulated array (TMA) is a simple antenna array architecture in which each antenna is connected to the multi-throw switch that performs time modulation~\cite{tma_survey}. The idea was first proposed in~\cite{shanks_tma_concept}. To improve the efficiency of time modulation, the switch can be replaced with a low-resolution discrete phase shifter~\cite{lin_phase_approx_tma_eff_analysis}. The simplicity of the switch, which is the primary building block of the front end, offers savings in cost, power, and size of the radio front end compared to mixers or high-resolution phase shifters~\cite{mmw_tech_book, rfic_book}.

However, the simple architecture of TMA comes with a few limitations that prevent its widespread adoption. Firstly, in conventional time-modulated arrays, the switch must operate at a multiple of the sampling frequency to take advantage of pulse oversampling to offer an improved phase-shifting resolution and prevent the overlap of the harmonics of the transmitted signal \cite{tma_survey, overview_tma_fda, mw_beamforming_w_tmas}. Secondly, due to the discretized modulating sequence, the time modulation introduces harmonic replicas of non-negligible power of the baseband signal at multiples of the switching frequencies. \cite{lin_phase_approx_tma_eff_analysis, signal_radiation_and_losses_in_tma}. To comply with wireless standards, the sideband radiation (SR) must be suppressed, which requires an additional bandpass filter per antenna. 
To facilitate more widespread utilization of TMAs, methods are needed to reduce the switching frequency and suppress the sideband radiation.
The harmonic components might be exploited to facilitate the simultaneous multi-beam operation with TMA \cite{harmonic_beamforming_tma, high_efficiency_switched_tma}. However, each harmonic beam transmits the same replicated baseband signal, limiting the practical application of transmit TMAs.

\begin{figure}[t]
    \centering
    \includegraphics[width=\linewidth]{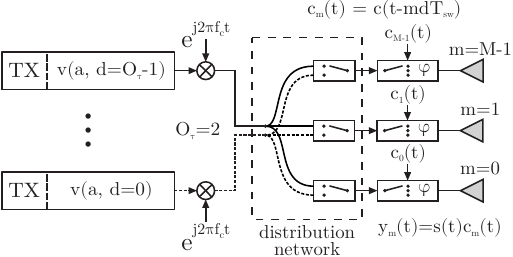}
    \caption{Architecture of the aliased time-modulated (ATMA) transmit array. The baseband signal consists of $A$ identical repeated blocks in the frequency domain. Each block is differently precoded by $v(a, d)$, depending on the block index $a$ and selected delay value $d$. The pulse oversampling factor $O_{\tau}$ determines the number of required frequency precoders implemented as digital baseband units. Depending on the index $d$, each $m$-th antenna is fed by a selected precoder via a distribution network of standard radio frequency (RF) switches.}
    \label{fig:atma_diagram}
\end{figure}

\subsection{Relevant works}

\begin{table*}[tb]
    \caption{State-of-the-art comparison}
    \label{tab:sota_comparison}
    \centering
    \def\arraystretch{1.2}
    \begin{tabular}{ 
    >{\centering\arraybackslash}m{0.075\linewidth}<{}|>
    {\centering\arraybackslash}m{0.15\linewidth}<{}|>
    {\centering\arraybackslash}m{0.1\linewidth}<{}|>
    {\centering\arraybackslash}m{0.075\linewidth}<{}|>
    {\centering\arraybackslash}m{0.075\linewidth}<{}|>
    {\centering\arraybackslash}m{0.125\linewidth}<{}|>
    {\centering\arraybackslash}m{0.075\linewidth}<{}
    }
         Reference & Modulation & Max measured SR [dB] & $f_{\mathrm{sw}}$ [MHz] & $B$ [MHz] & Effective number of phase shifts & Optimization required \\
         \hline
         \cite{sb_supp_nonuniform} & Binary amplitude & $-$35.98 & 1.9 & 1 & N/A & No \\ 
         \hline
         \cite{sb_suppresion_tma_freq} & Binary amplitude & $-$36.94 & 0.45 & 0.1 & N/A & Yes \\  
         \hline
         \cite{tma_cont_spectrum} & Phase and binary amplitude & $-$39.4 & 0.5 & N/A & $\infty$ & No \\
         \hline
         \cite{sb_supp_stair_pulses} & IQ (256) & $-$33.3 & 25.6 & 0.4 & 256 & Yes \\
         \hline
         \cite{sb_supp_for_wide_bandwidth} & IQ (16) & $-$33.57 & $\infty$ & N/A & $\infty$ & Yes \\
         \hline
         \cite{iq_ssb_tma} & IQ (4) & $-$13.98 & $\infty$ & 0.4 & $\infty$ &  No \\
         \hline
         \cite{iq_modulation_enhanced_tma} & IQ (8) & $-$16.9 & 0.8 & 0.8 & 8 & No \\
         \hline
         This work & Phase & $-$6.57 \footnotemark & $0.625$ & 20 & $8$ & No \\
    \end{tabular}
\end{table*}

Initially, time-modulated arrays were analyzed under a narrowband scenario, considering just a single carrier. Various algorithms have been proposed to optimise the switching instants of the RF switches to minimise sideband radiation, including genetic \cite{tma_w_discrete_slots_and_ga}, evolution \cite{sb_supp_evol_alg, phase_modulation_delay_line_tma} and particle swarm \cite{tma_stability_opt_particle_swarm} optimization algorithms.
The modulating signal format determines the sideband radiation. To further reduce sideband radiation, the following approaches have been proposed: aperiodic modulating pulses \cite{sb_supp_nonuniform}, pseudorandom modulating sequence \cite{tma_cont_spectrum} or different switching frequency per element \cite{sb_suppresion_tma_freq}. A significant improvement in the sideband suppression is observed when employing phase modulation \cite{iq_ssb_tma}, utilizing stepped waveforms \cite{sb_supp_stair_pulses} or simultaneous amplitude and phase modulation (IQ) \cite{sb_supp_stair_pulses, sb_supp_for_wide_bandwidth, iq_modulation_enhanced_tma}. The interest in the TMA architecture has resulted in practical implementations which utilize simple low-bit switched delay lines as building blocks \cite{2bit_phase_modulation_tma, 2bit_high_res_phase_mod_time_delay}. On the other hand, the interference from the aliases of the signal can be deliberately exploited to generate artificial noise and enable directional modulation, which is achieved at the cost of beamforming gain \cite{tma_ofdm_directional, tma_ofdm_security, switched_pa_security, tma_irs_gen}. However, the out-of-band radiation generated by the harmonic replicas of the signal in \cite{tma_ofdm_directional, tma_ofdm_security, tma_irs_gen} is not addressed and often assumed to be filtered out or negligible.

The listed approaches are suitable for narrowband signals, and the switching frequency is assumed to be much higher than the transmitted signal bandwidth. This leads to an assumption of an arbitrary and unconstrained timing resolution of a switch. 
For wideband signals, this assumption is no longer valid and the switching frequency becomes constrained by signal bandwidth and quickly approaches limits imposed by hardware. Moreover, the extent of the sideband radiation is increased as the baseband signal now has a greater width in the frequency domain, increasing the strain on filtering requirements.

\footnotetext{Based on the minimal working example with $N=2$ and $A=128$. The aliasing improves the SR over a wide bandwidth by averaging oppositely precoded blocks. However, the power of the nearest blocks remains relatively high, due to an imbalance in the resulting phase of the aliased (summed) blocks. When considering a transition bandwidth of 1\%, the maximum measured SR is reduced to $-$18.66 dB and $-$36.32 dB for 10\%. See the experimental results in Sec. \ref{sec:experimental_validation}. }

\subsection{Contributions}

This work considers the TMA combined with a wideband OFDM signal and proposes a method to simultaneously reduce the switching frequency and the sideband radiation by introducing aliasing between the harmonic replicas of the transmitted signal. It is achieved at the cost of the effective bandwidth as the baseband signal, in the frequency domain, is composed of $A$ repeated blocks, which are differently precoded. The improvement in sideband suppression and spectral efficiency trade-off is discussed in detail. The pulse oversampling of the modulating sequence for improved resolution beamforming with aliased TMA (ATMA) is proposed and studied. The feasibility of the proposed method is validated experimentally for wideband signals.
Based on the experimental measurements of the time modulator, the full-wave array simulation is performed to validate the beamforming performance.

Table \ref{tab:sota_comparison} presents the work with reference to the state-of-the-art. The fields with N/A denote that the authors did not consider the parameter in the analysis or did not specify it directly in the measurement setup. The fields with $\infty$, especially at switching frequency $f_{\mathrm{sw}}$ and effective number of phase shifts, indicate that the authors assume arbitrary switching instants and the resultant phase shifting resolution is arbitrarily high. Note that binary amplitude modulation has low efficiency due to the switching off of the transmitted signal. The IQ modulation requires complex hardware and to achieve low SR values, optimization is required, which is cumbersome in real-time applications. All of the reported works require a switching frequency higher than the signal bandwidth. The proposed solution offers a low complexity and scalable method to simultaneously reduce sideband radiation and switching frequency. In the proposed method, the maximum measured SR is primarily determined by the phase shifter resolution. The aliasing still improves the maximum SR compared to a system without it; however, its primary benefit is the reduction of the sideband radiation over a wide bandwidth. When some transition bandwidth is considered, the maximum SR is significantly reduced due to the steep spectral roll-off of the aliased OFDM spectra.

This article is organized as follows. Section \ref{sec:sig_mod} presents the principles of the time modulation and the signal model. The aliasing is discussed in Section \ref{sec:aliasing} along with the key performance improvement metrics. Finally, the scheme is validated in Section \ref{sec:experimental_validation}. Finally, Section \ref{sec:conclusion} concludes the paper.  

\section{Signal model}
\label{sec:sig_mod}

\subsection{System design}
This work proposes a time-modulated OFDM transmit system that supports wideband signals. The system is composed of multiple components, which, combined, offer a simple and cost-effective solution for analog beamforming. 
First, Sec. \ref{sec:time_modulation} discusses the time modulation with a discrete phase shifter and the harmonic components generated by switching. Next, the phase shifting of the baseband signal is discussed in Sec. \ref{sec:phase_shifting}, which is achieved by the cyclic delay of the modulating sequence. In this work, the switching frequency is lower than the signal bandwidth, resulting in aliasing of the harmonic copies of the baseband signal. To make the transmitted signal robust against aliasing repeated subcarrier block structure is proposed in Sec. \ref{sec:aliasing}.
Next, to reconstruct the original signal after the time-modulation and reduce the out-of-band radiation, the coherent combining of the aliases needs to be controlled by the proper precoder design, which is discussed in Sec. \ref{sec:prec_design}. Finally, the beamforming performance of the ATMA is discussed in \ref{sec:beamforming_w_tma}. The constraints on the choice of the system parameters are mentioned in \ref{sec:sys_consideration}.

\subsection{Time modulation}
\label{sec:time_modulation}
Consider a time-modulated antenna, which is connected to the $ N$-throw/state switch. The RF switch acts as a discrete phase shifter - each state of the switch results in a different phase shift of the input signal.
The phase shift of the $n$-th switch state is given by
\begin{equation}
    \label{eq:phase_shift_per_idx}
    \varphi(n) = 2 \pi \frac{n}{N},
\end{equation}
where $n$ is the index of the state $n \in \{0, 1, \ldots, N-1 \}$.
Consider a bandlimited transmit signal of bandwidth $B$ and corresponding sampling frequency $f_\mathrm{s} = B$.
While time modulation is fundamentally an analog-domain operation, practical wideband systems require synchronization between the digital waveform generation and the analog switching sequences. For wideband signals, the symbol duration becomes comparable to the time-modulation switching period. Therefore, the signal bandwidth expressed in terms of sample rate $f_{\mathrm{s}}$ is the central parameter of the considered system and analysis.

The discrete phase shifter supports the switching frequency $f_{\mathrm{sw}}$, which is a fraction of the sampling frequency given by
\begin{align}
    \label{eq:sw_freq}
    f_{\mathrm{sw}} = f_\mathrm{s} \frac{O_{\tau}}{A}, \quad A,O_{\tau} \in \mathbb{N}^+,
\end{align}
where $A$ is the aliasing factor that controls the pulse duration. The $O_{\tau}$ is the pulse oversampling factor that does not affect the pulse duration but controls the number of sampling instances within the pulse, allowing for finer control of the pulse delay. 

In the following, the harmonic components of the modulating sequence are derived; the pulse oversampling is omitted from the analysis, as it is only used to control the delay of the modulating sequence. The impact of pulse oversampling is discussed in the following Sec. \ref{sec:phase_shifting}.
The pulse duration of each phase state is a multiple of the sampling period $T_{\mathrm{p}} = A T_{\mathrm{s}}$ and the resulting pulse frequency is a fraction of the sampling frequency
\begin{align}
    \label{eq:pulse_freq}
    f_{\mathrm{p}} = \frac{1}{T_{\mathrm{p}}} = \frac{f_{\mathrm{s}}}{A}.
\end{align}

\begin{figure}[b]
    \centering
    \includegraphics[width=\linewidth]{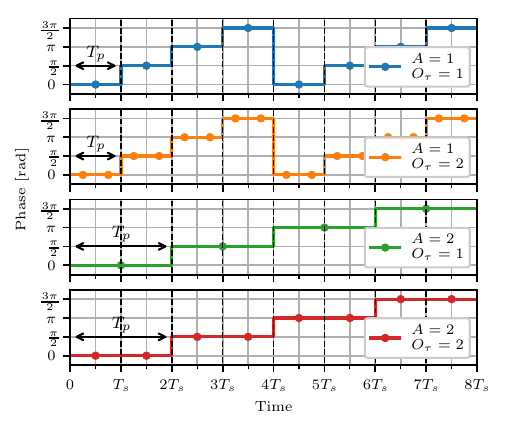}
    \caption{Modulating signal of the ATMA for $N=4$ and different aliasing $A$ and pulse oversampling $O_{\tau}$ factors.}
    \label{fig:atma_sig_td}
\end{figure}
The modulating signal is a periodically repeated sequence of $N$ rectangular pulses with a phase given by \eqref{eq:phase_shift_per_idx}. This corresponds to discrete linear phase modulation that maximizes the power at the first harmonic \cite{lin_phase_approx_tma_eff_analysis}.
Due to the discretized switching instants and phase shift values, the basic building block of the TMA modulating signal is a rectangular pulse. Consider a periodic rectangular pulse of duration $T_{\mathrm{p}}$ with unit amplitude and phase shift of $\varphi(n)$.  The time domain representation of the pulse is given by 
\begin{equation}
    \label{eq:td_rect_pulse}
    g_n(t) = e^{j\varphi(n)}\sum_{i=-\infty}^{\infty}\rect{\left( \frac{t}{T_\mathrm{p}} - \frac{T_{\mathrm{p}}}{2} + i N T_{\mathrm{p}} \right)}.
\end{equation}
Fig. \ref{fig:atma_sig_td} illustrates the modulating signal in the time domain.
The pulse duration is determined by the aliasing factor $A$, while the pulse oversampling factor only affects the sampling granularity. The finer sampling grid enables finer cyclic delay of the modulating sequence, which is used to control the phase shift of the modulated signal.
The pulse is periodic with a period $N T_{\mathrm{p}}$, which allows expanding it into a Fourier series.
The Fourier coefficient of the $k$-th harmonic component is given by
\begin{align}
    \label{eq:fourier_coeffs}
    G_n(k) &= e^{j\varphi(n)} \sinc{\left(\pi \frac{k}{N} \right)} e^{-j\pi \frac{k}{N}},
\end{align}
where $\sinc{(x)} = \sin{x} / x$.
The modulating (control) signal $c(t)$ is composed of a sequence of $N$ delayed complex pulses from \eqref{eq:td_rect_pulse} resulting in
\begin{align}
    \label{eq:pulse_seq_td}
    c(t) &= \sum_{n=0}^{N-1} g_n \left(t - n T_{\mathrm{p}} \right).
\end{align}
The frequency domain representation of the modulating signal is
\begin{align}
    \label{eq:fourier_coeffs_pulse_seq}
    C(k) &= \frac{1}{N} \sum_{n=0}^{N-1} G_n(k) e^{-j2\pi \frac{n}{N} T_{\mathrm{p}} f_{\mathrm{p}} } \nonumber \\
    &= \sinc{\left(\pi \frac{k}{N} \right)} e^{-j\pi \frac{k}{N}} I(k)
\end{align}
$I(k) = \frac{1}{N} \sum_{n=0}^{N-1} e^{j2\pi \frac{n}{N} \left( 1 - k \right)}$ determines the existence of the $k$-th spectral component at frequency $f_k = \frac{k}{N} f_{\mathrm{p}}$
\begin{align}
    \label{eq:fourier_comp_existence_discr_freq}
    I(k) &= \begin{cases} 
        1, & k = 1 + zN \\ 
        0, & k \neq 1 + zN
    \end{cases},
    \quad z \in \mathbb{Z}
\end{align}

To simplify the following analysis, the discrete harmonic representation of the modulating sequence is transformed into a continuous frequency range
\begin{align}
    \label{eq:mod_seq_fd_continious}
     C(f) &= \alpha(i) \delta \left( f - \left(\frac{f_{\mathrm{p}}}{N} + i f_{\mathrm{p}} \right) \right)
\end{align}
where $\alpha(i) = \sinc{\left(\pi \left( i + \frac{1}{N} \right) \right)} e^{-j\pi \left( i + \frac{1}{N} \right) }$ is the complex coefficient determining the amplitude and phase of the $i$-th harmonic component and $\delta$ is the Dirac delta function. The $i$-th harmonic components are located at frequency $\frac{f_{\mathrm{p}}}{N} + i f_{\mathrm{p}}$.
To gain more insight into the amplitude and phase of the harmonic components, $\alpha(i)$ can be expressed as
\begin{align}
    \label{eq:h_ampl_phase}
    \alpha(i) 
    =& \left| \sinc{\left(\pi \left(i + \frac{1}{N}  \right) \right)} \right| e^{j \phi_{\alpha}(i)},
\end{align}
where $\phi$ is the function describing the phase of the $\alpha(i)$ harmonic component
\begin{equation}
    \label{eq:phase_per_harm}
    \phi_{\alpha}{\left(i \right)} = \begin{cases} 
        -\frac{\pi}{N}, & i \geq 0 \\ 
        -\frac{\pi}{N} - \pi, & i < 0
    \end{cases}.
\end{equation} 

\begin{figure}[b]
    \centering
    \includegraphics[width=\linewidth]{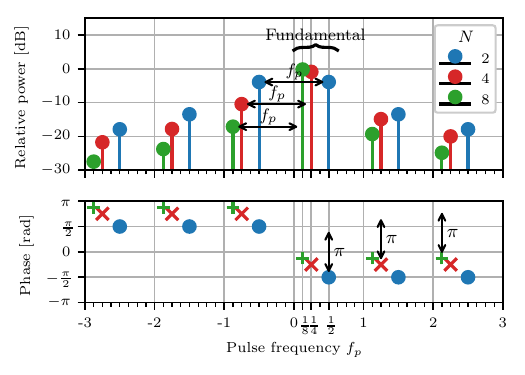}
    \caption{Power and phase of the harmonic components of the time modulating signal for $A=1$ and selected values of $N$.}
    \label{fig:tma_noalias_h_mag_phase_vs_nphase}
\end{figure}

The length of the modulating sequence $N$ affects the power of the harmonics by narrowing the main lobe of the $\sinc$ function. The frequency shift of the total signal is equal to $f_{\mathrm{mod}} = f_{\mathrm{p}} / N$ - the modulating frequency. The length $N$ affects the phase of the harmonic components; however, note that the relative difference between the positive and negative components is constant and equal to $\pi$. The aliasing factor $A$ reduces the spacing of harmonic components and the total frequency shift of the transmitted signal. Fig. \ref{fig:tma_noalias_h_mag_phase_vs_nphase} shows the power and phase of the harmonics for $A=1$ and selected values of $N$.

\subsection{Phase shifting}
\label{sec:phase_shifting}
The phase shifting of the signal is achieved by a cyclic shift of the modulating sequence.
The pulse oversampling allows for obtaining a refined timing resolution, which results in a multiplied number of effective phase shifts.
Note that in order to have a switching frequency lower than the signal bandwidth, the pulse oversampling factor should always be lower than the aliasing factor, $O_{\tau} \leq A$ \eqref{eq:sw_freq}.
Given a sequence of length $N$ and pulse duration $T_{\mathrm{p}}$, there are $D$ possible discrete delay values given by
\begin{equation}
    \label{eq:num_phase_shifts}
    D = N \frac{T_{\mathrm{p}}}{T_{\mathrm{sw}}} = N \frac{T_{\mathrm{p}}}{\frac{T_{\mathrm{p}}}{O_{\tau}}} = NO_{\tau},
\end{equation}
where $T_{\mathrm{sw}} = 1/f_{\mathrm{sw}}$.
Note that, despite the pulse frequency being $f_{\mathrm{s}}/A$, the switch has to support an $O_{\tau}$ higher switching frequency to be able to dynamically control the delay between subsequent transmitted OFDM symbols and modulating sequences. 
An example of an oversampled modulating signal is shown in Fig. \ref{fig:atma_sig_td}.

The delay (cyclic shift) of the time domain modulating sequence is equivalent to the phase rotation in the frequency domain as follows
\begin{align}
    \label{eq:time_delay_mod_sig_short}
    C(f, d) &= \mathcal{F}\left\{ c \left( t - d T_{\mathrm{sw}} \right) \right\} \nonumber \\
    &=\alpha(i) e^{-j2\pi d T_{\mathrm{sw}} f} \delta{\left( f - \frac{f_{\mathrm{p}}}{N} - i f_{\mathrm{p}} \right)},
\end{align}
where $d \in \{0,1, \ldots, D-1\}$ is the index of the selected digital delay.
The phase shift due to the cyclic delay of the phase modulating sequence is
\begin{align}
    \label{eq:ph_shift_per_harm_per_delay}
    \varphi_{\tau}(i, d) &= -2\pi d T_{\mathrm{sw}} \left( \frac{f_{\mathrm{p}}}{N} + i f_{\mathrm{p}} \right) \nonumber \\ 
    &= -2\pi \frac{d}{D} - 2\pi \frac{d}{O_{\tau}} i.
\end{align}
It can be seen that the pulse oversampling increases the phase-shifting resolution by an order of $O_{\tau}$. However, for $d$ values other than multiples of $O_{\tau}$, there is an additional phase shift dependent on the harmonic index $i$.
Extending \eqref{eq:h_ampl_phase} with \eqref{eq:ph_shift_per_harm_per_delay} to include the effects of the time delay on the harmonic component results in
\begin{equation}
    \label{eq:ph_shift_i_d_sep}
    \alpha(i, d) = \alpha(i) e^{j \varphi_{\tau}(i, d)}.
\end{equation}

\subsection{Aliasing}
\label{sec:aliasing}

The maximum bandwidth of the signal transmitted by the conventional TMA to avoid aliasing is determined by the spacing of the harmonic components equal to $f_{\mathrm{s}} / A$ \eqref{eq:pulse_freq}. In this paper, aliasing combined with the block structure of the signal in the frequency domain and precoding is exploited to introduce destructive interference between the harmonics of the transmitted signal to reduce the sideband radiation.
Consider a baseband signal of bandwidth $B = f_{\mathrm{s}}$ that consists of $A$ (aliasing factor) joint/adjacent rectangular blocks of equal width $B/A$ in the frequency domain. The frequency domain representation of the single block is
\begin{equation}
    \label{eq:bb_sublock}
    S_a(f) = v(a) \rect{\left( \frac{f}{\frac{f_{\mathrm{s}}}{A}} \right)},
\end{equation}
 where $v(a)$ is complex frequency-flat precoding coefficient per block and $a \in \{ 0, 1, \ldots, A-1 \}$ is the block index.
The total baseband signal can be written as a convolution of a precoded finite Dirac delta comb of length $A$ and a rectangular function
\begin{align}
    \label{eq:total_bb_signal}
    S(f) &= \sum_{a = 0}^{A - 1}  S_a\left(f - a \frac{f_{\mathrm{s}}}{A} + \frac{A - 1}{2A} f_{\mathrm{s}} \right) \\
    &= \rect{\left( \frac{f}{\frac{f_{\mathrm{s}}}{A}} \right)} * \sum_{a = 0}^{A - 1} v(a) \delta{\left( f - \left(a - \frac{A-1}{2}\right) \frac{f_{\mathrm{s}}}{A}\right)} \nonumber.
\end{align}

In this system, unlike conventional time-modulated arrays, the baseband signal is composed of repeated subcarrier blocks. Applying time modulation with a fraction of the signal bandwidth introduces aliasing and coherent combination between the subcarrier blocks. After the time modulation, the originally transmitted signal of bandwidth $B$ is recreated by the coherent combination of the harmonic components. The repeated subcarrier blocks with appropriate precoding allows for an effective reconstruction of the signal after the time modulation. The time modulation of the considered signal offers a low-cost, simple phase shifting method at the cost of repeated block structure and generating harmonic components outside of the initially specified band $B$.

The time modulation of this signal with a precoded-shifted frequency-domain representation is achieved by multiplying the time domain signal with a modulating sequence $C(f, d)$, which in the frequency domain corresponds to the convolution of \eqref{eq:mod_seq_fd_continious}  with \eqref{eq:total_bb_signal} 
\begin{align}
    \label{eq:tma_bb_conv}
    &Y(f, d) = S(f) * C(f, d) \nonumber \\
    &\quad = \sum_{a = 0}^{A - 1} v(a) \alpha(i, d) \delta{\left( f - \left( a -\left\lfloor\frac{A-1}{2}\right\rfloor + \Delta_{f} + i \right) \frac{f_{\mathrm{s}}}{A} \right)} \nonumber \\ 
    &\qquad * \rect{\left( \frac{f}{\frac{f_{\mathrm{s}}}{A}} \right)},
\end{align}
where $\Delta_{f} = \left\lfloor\frac{A-1}{2}\right\rfloor - \frac{A-1}{2} + \frac{1}{N}$ is a fractional shift in frequency due to time modulation and centering of the convolution kernel $v(a)$.
The aliased signal with overlapping frequency blocks can be written as
\begin{align}
    \label{eq:tma_bb_conv_aliased}
    Y(f, d) =& \sum_{a = 0}^{A - 1} v\left(a \right) \alpha \left(i - a + \left\lfloor \frac{A-1}{2} \right\rfloor, d \right) \nonumber \\ 
    & \rect{\left( \frac{f}{\frac{f_{\mathrm{s}}}{A}} \right)} * \delta{\left( f - \left( \Delta_{f} + i \right) \frac{f_{\mathrm{s}}}{A} \right)} \nonumber \\
    =& \alpha_{\mathrm{A}}(i, d) \rect{\left( \frac{f}{\frac{f_{\mathrm{s}}}{A}} - \left( \Delta_{f} + i \right) \right)}.
\end{align}
where $\alpha_{\mathrm{A}}(i, d)$ is the complex coefficient resulting from aliasing of $A$ precoded harmonic components $\alpha(i, d)$. Expanding $\alpha(i,d)$ with \eqref{eq:ph_shift_i_d_sep} and \eqref{eq:ph_shift_per_harm_per_delay} gives
\begin{align}
    \label{eq:aliased_h_coef}
    \alpha_{\mathrm{A}}(i, d) =& e^{-j2\pi \frac{d}{D}} e^{-j2\pi \frac{d}{O_{\tau}} i} e^{-j2\pi \frac{d}{O_{\tau}} \left\lfloor \frac{A-1}{2} \right\rfloor} \nonumber \\
    & \sum_{a = 0}^{A - 1} v(a) e^{j2 \pi \frac{d}{O_{\tau}} a} \alpha \left( i - a + \left\lfloor \frac{A-1}{2} \right\rfloor \right)
\end{align}

\subsection{Precoder design}
\label{sec:prec_design}
The destructive interference of the sideband radiation is achieved by the design of the precoding vector $v(a)$. However, pulse oversampling introduces an additional phase component $e^{j2\pi a\frac{d}{O_{\tau}}}$ in the sum of \eqref{eq:aliased_h_coef}. To compensate for the phase shift, the precoder is extended with a conjugate of it
\begin{equation}
    \label{eq:prec_extended}
    v(a, d) = v(a) e^{-j 2 \pi a \frac{d}{O_{\tau}}}.
\end{equation}
The pulse oversampling $O_{\tau}$ determines the periodicity of the phase shift and the number of additional precoding vectors required to compensate for it across all $d$ values. The precoder is chosen based on the $d$ value per antenna. 
When considering pulse oversampling, multiple precoders are required, increasing the complexity of the array architecture. As digital precoding is required, the precoders are implemented as separate digital basebands. Each antenna might require a different precoder, which can be provided by a switched distribution network, as shown in Fig. 1. 
The baseband complexity (number of digital baseband units) is proportional to the oversampling factor $O_{\tau}$. 
For a system with $M$ antennas, each time modulator and antenna has to be provided with a choice of precoder which requires $O_{\tau}$ input and $M$ output distribution network built with $O_{\tau}$, $M$-way splitters and $M$ single-pole $O_{\tau}$-throw RF switches that route the selected precoder to the time-modulating switch and antenna. The switching network complexity scales proportionally to $M O_{\tau}$.

Given the extended precoder the sum in \eqref{eq:aliased_h_coef} simplifies to 
\begin{align}
    \label{eq:simpl_aliased_h_coef}
    \alpha_{\mathrm{A}}(i, d) =& e^{j \varphi_{\tau_{\mathrm{A}}}(i,d)} \sum_{a = 0}^{A - 1} v(a, d) \alpha \left( i - a + \left\lfloor \frac{A-1}{2} \right\rfloor \right),
\end{align}
where $e^{j \varphi_{\tau_{\mathrm{A}}}} = e^{-j2\pi \frac{d}{D}} e^{-j2\pi \frac{d}{O_{\tau}} \left(i + \left\lfloor \frac{A-1}{2} \right\rfloor \right)}$ is the phase shift due to the selected delay value $d$.

Given that the initial transmitted baseband signal was limited to $-B/2$ to $B/2$, the harmonic indices that lie within this band are referred to as the desired. While the harmonic indices outside of this frequency band are referred to as the undesired harmonic components.
To reconstruct the original baseband signal, the interference between the aliases within the desired signal indices should be constructive, while it should be destructive for the others to minimize the sideband radiation.
The desired baseband signal is within the harmonic indices $i_{\mathrm{bb}} \in \{ -(A-1) + \left\lfloor \frac{A-1}{2} \right\rfloor, \ldots, \left\lfloor \frac{A-1}{2} \right\rfloor \}$. All other harmonic components $i \notin i_{\mathrm{bb}}$ are considered sideband radiation and should be minimized.
By evaluating the phase of the harmonic components for $ i > \left\lfloor \frac{A-1}{2} \right\rfloor$ it can be seen that the argument of $\alpha$ in \eqref{eq:simpl_aliased_h_coef} is always positive, resulting in constant phase given by \eqref{eq:phase_per_harm} for all positive sideband harmonic indices. Similarly for the negative sideband indices $ i < -(A-1) + \left\lfloor \frac{A-1}{2} \right\rfloor$ the argument of $\alpha$ is always negative. Therefore, the harmonics on the positive and negative sides of the sideband have a constant phase within each sideband.
The amplitude of the harmonic components decreases with increasing $|i|$. Each sideband exhibits a different monotonicity and rate of change/slope (for $N>2$) of the amplitude of the harmonic components. This renders the amplitude precoding cumbersome and limits its feasibility. 

In the following, a phase-only precoder is considered to minimize both sidebands simultaneously. 
Given the constant phase of the summed sideband harmonic components, a natural choice to minimize the sum is to alter the sign of the summed components by applying an alternating precoder
\begin{equation}
    \label{eq:alternating_precoder}
    v(a) = (-1)^a = e^{j \pi a} .
\end{equation}
The alternating precoder introduces destructive interference between the harmonic components at the sidebands at the cost of introducing power variations in the power of the passband signal. The fluctuations result from the summation of a different subset of precoded harmonic components. The $0$-th harmonic component dominates the amplitude of the result and determines the phase.
Expanding the precoder from \eqref{eq:prec_extended} with the alternating phase from \eqref{eq:alternating_precoder} yields
\begin{align}
    \label{eq:prec_total}
    v(a, d) 
    &=e^{j \pi a \left(1 - \frac{2d}{O_{\tau}} \right)}.
\end{align}

Note that the aliased signal in the desired passband range has an alternating phase precoding, which can be easily compensated at the receiver. 
After the convolution/aliasing, indices of baseband signal of interest have a flipped range  $i_{\mathrm{bb}_{\mathrm{A}}} =  \{ -\left\lfloor \frac{A-1}{2} \right\rfloor, \ldots, A-1 - \left\lfloor \frac{A-1}{2} \right\rfloor \}$. 

\begin{figure}[t]
    \centering
    \includegraphics[width=\linewidth]{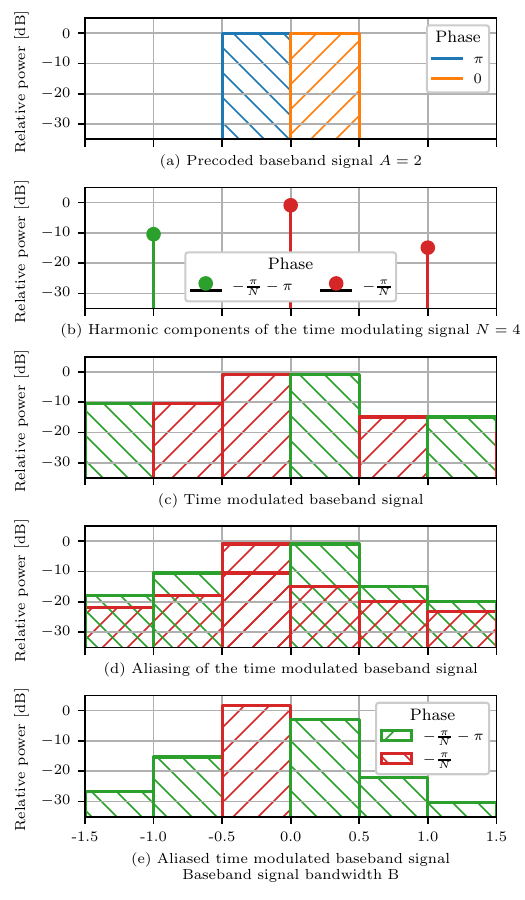}
    \caption{Illustration of the aliasing for $N=4$ and $A=2$, no pulse oversampling $O_{\tau}=1$. }
    \label{fig:atma_aliasing_exp}
\end{figure}

Fig. \ref{fig:atma_aliasing_exp} illustrates the signal processing steps in the aliased TMA system with $A=2$ blocks. Fig. \ref{fig:atma_aliasing_exp}a presents the precoded baseband signal \eqref{eq:total_bb_signal} in the frequency domain. Fig. \ref{fig:atma_aliasing_exp}b shows the harmonic components of the modulating signal from \eqref{eq:h_ampl_phase} with the color denoting the phase of each harmonic component. Next, the precoded baseband signal is multiplied by the modulating signal, which corresponds to convolution in the frequency domain, resulting in replication of the baseband spectrum with altered phase precoding as shown in Fig. \ref{fig:atma_aliasing_exp}c. Note how the precoder phase affects the resultant phase of the blocks after the time modulation. To emphasize the phases of the baseband blocks, before aliasing and hence the spacing of the harmonics, the modulating frequency is set to be equal to the baseband bandwidth $B$ in Fig. \ref{fig:atma_aliasing_exp}b and \ref{fig:atma_aliasing_exp}c. The aliasing of the differently precoded baseband blocks is illustrated in Fig. \ref{fig:atma_aliasing_exp}d according to \eqref{eq:tma_bb_conv_aliased}. The aliasing in Fig. \ref{fig:atma_aliasing_exp}d and \ref{fig:atma_aliasing_exp}e is achieved by reducing the modulating frequency by $2$, making it $B/2$. At the same time, the baseband bandwidth stays the same, resulting in an overlap of neighbouring blocks. Fig. \ref{fig:atma_aliasing_exp}d shows the phases of the aliased blocks before combining. It helps to observe where the noncoherent combination occurs and estimate the resulting power per block. Finally, Fig. \ref{fig:atma_aliasing_exp}e shows the sum of the aliased baseband blocks from Fig. \ref{fig:atma_aliasing_exp}d with visible attenuation of the sideband blocks. Note that the phase after the aliasing is not uniform across baseband blocks, which has to be taken into account at the receiver.

\begin{figure}[t]
    \centering
    \includegraphics[width=\linewidth]{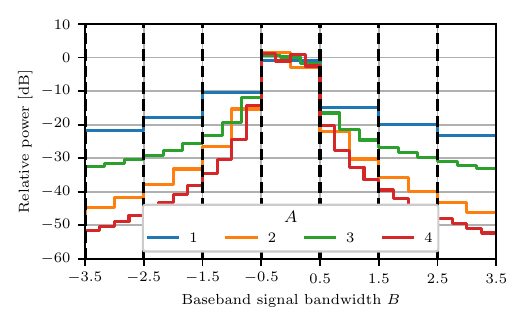}
    \caption{Spectrum of TMA with considered block signal in frequency domain for $N=4$ and selected values of $A$. (The frequency shift due to modulation is neglected.)}
    \label{fig:atma_spectrum}
\end{figure}

Fig. \ref{fig:atma_spectrum} shows the signal power spectrum at the time-modulator output for $N=4$ and selected values of the aliasing factor $A$. The aliasing with an alternating precoder effectively reduces the power of the sideband radiation while introducing slight variations (ripple) in the power of the transmitted signal in the center of the spectrum. The sideband radiation cancellation is less effective for an odd $A$ - an odd number of aliasing terms due to a lack of balance in the number of summed positively and negatively precoded components. On the contrary, odd $A$ factors increase the spectral roll-off and reduce the requirements on the transition bandwidth.
The practicality of odd $A$ factors is limited by the fact that the number of subcarriers $K$ has to be a multiple of $A$, which might result in an odd number of subcarriers, presenting implementation challenges of the FFT and reducing compatibility with standards and deployed systems. 
Due to the poorer performance of the aliased system with odd values of $A$, the remaining results consider only even values of $A$; nevertheless, the signal model and discussion are valid for any $A$. 

\begin{figure}[t]
    \centering
    \includegraphics[width=\linewidth]{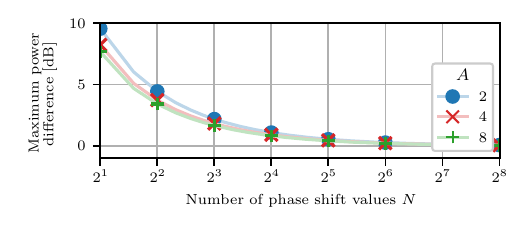}
    \caption{The maximum power difference between blocks in the passband due to aliasing for selected values of $A$.}
    \label{fig:pwr_diff_ampl_vs_nphase_steps}
\end{figure}

The peak-to-peak power variations per block are determined mainly by $N$, and increasing $A$ has a limited impact as the amplitude difference of the adjacent harmonics determines the summation result. Increasing $N$ improves the power of the main harmonic while reducing the power of other, reducing the ripple. Fig. \ref{fig:pwr_diff_ampl_vs_nphase_steps} presents the maximum power difference between the blocks in the frequency domain within the band of interest as a function of $N$. The power variations become negligible for $N \geq 16$.

The proposed system offers two ways of reducing the sideband power. The first is by increasing the number of switch states and phase shifts $N$, which narrows the sinc lobe, see \eqref{eq:fourier_coeffs}. The second is by increasing the aliasing factor $A$. To quantify the reduction in the sideband power, the adjacent channel leakage ratio (ACLR) is used as a figure of merit. The ACLR is defined as
\begin{equation}
    \label{eq:aclr_def}
    \mathrm{ACLR} =  \frac{\sum_{i \in i_{\mathrm{bb}_{\mathrm{A}}} } |\alpha_{\mathrm{A}}(i)|^2 } {\sum_{i \in (i_{\mathrm{bb}_{\mathrm{A}}} - A)} |\alpha_{\mathrm{A}}(i)|^2 }.
\end{equation}

\begin{figure}[t]
    \centering
    \includegraphics[width=\linewidth]{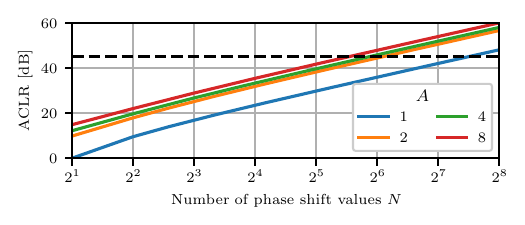}
    \caption{ACLR versus the number of phase shift values $N$ for selected number of blocks $A$.}
    \label{fig:aclr_vs_n_phase_shift}
\end{figure}
\begin{figure}[t!]
    \centering
    \includegraphics[width=\linewidth]{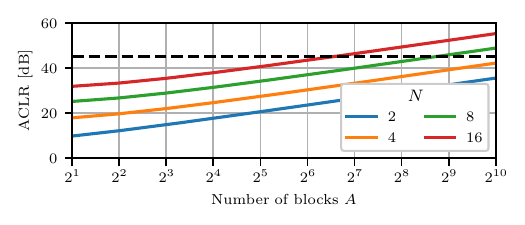}
    \caption{ACLR versus the number of blocks $A$ for selected number of phase shift values $N$.}
    \label{fig:aclr_vs_n_blocks}
\end{figure}
\begin{figure}[t!]
    \centering
    \includegraphics[width=\linewidth]{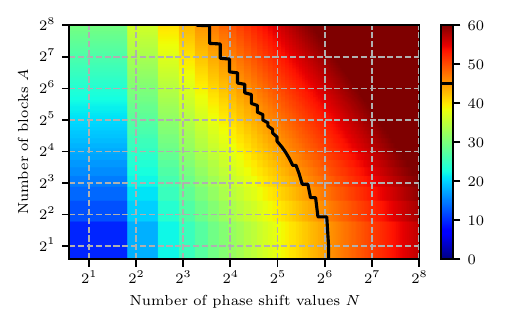}
    \caption{TMA ACLR as a function of number of blocks $A$ and the number of the phase shifter steps $N$. The black line denotes the 5G minimum requirement of 45 dB \cite{3gpp_nr}.}
    \label{fig:aclr_vs_phres_vs_nblocks}
\end{figure}
Fig. \ref{fig:aclr_vs_n_phase_shift} presents the ACLR for a fixed number of blocks $A$ for a varying number of phase shifts $N$. The dashed black line denotes the 5G minimum requirement of 45dB \cite{3gpp_nr}. The introduction of aliasing improves the ACLR by around 7 dB compared to a system without it. Each doubling of $N$ improves the ACLR by around 6.1 dB at the cost of increased hardware complexity. 
Note that the ACLR is defined as the ratio of the integrated power within the main channel bandwidth to the integrated power in the adjacent channel. It can be observed in Fig. \ref{fig:atma_spectrum} that the maximum sideband radiation is not significantly affected by the aliasing factor $A$. However, increasing $A$ steepens the spectral decay, reducing the transition bandwidth needed to reach a required low sideband radiation level. To achieve low values of the SR, the proposed method requires a transition region or guard bandwidth. 
The achievable SR values are dependent on the assumed guard bandwidth and should be interpreted together with the specified guard bandwidth.

Fig. \ref{fig:aclr_vs_n_blocks} shows the ACLR for fixed values of $N$ and a varying number of blocks $A$. For a small number of blocks, the ACLR gains are minor; however, as $A$ increases, each doubling of $A$ improves the ACLR by around 2.9 dB. Aliasing reduces the power of the sideband radiation without extensive hardware modifications.
As shown by Fig. \ref{fig:aclr_vs_n_phase_shift} and \ref{fig:aclr_vs_n_blocks}, satisfying the minimum ACLR for small values $A$ or $N$ requires the other parameter to be significantly large, which is not feasible. To meet the requirements, a reasonable approach is to optimize both parameters simultaneously. 
Fig. \ref{fig:aclr_vs_phres_vs_nblocks} presents the ACLR as a function of $N$ and $A$. The black line denotes the 5G requirement of a minimum of 45 dB \cite{3gpp_nr}. By tuning both parameters, an acceptable ACLR performance is reached for $N$ and $A$ equal to $2^5$ instead of requiring $N=2^7$ when $A=2$.

\subsection{Beamforming with aliased TMA}
\label{sec:beamforming_w_tma}
\begin{figure}[b]
    \centering
    \includegraphics[width=\linewidth]{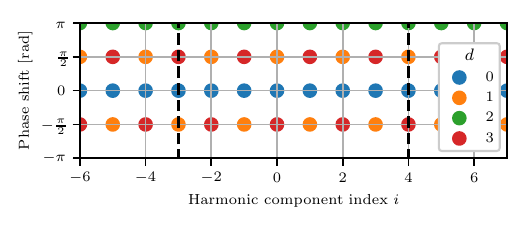}
    \caption{Phase shift of harmonic components for different values of $d$ normalized with regard to $d=0$. For $A=8$, blocks, $N=2$, and $O_{\tau}=2$. The black dashed lines denote the range of the desired baseband indices $i \in i_{\mathrm{bb}_\mathrm{A}}$. }
    \label{fig:ph_shift_per_d}
\end{figure}

The ULA TMA array is considered to be operating in the far field with isotropic antenna elements spaced by $d_{\lambda}$. Taking the first element as a reference, the path difference per antenna element in the direction $\theta$ is
\begin{align}
    \label{eq:path_diff_ped_ant}
    \Delta d_m &= m d_{\lambda} \frac{c}{f_{\mathrm{c}}} \sin{(\theta)},
\end{align}
where $d_{\lambda}$ is the spacing between antenna elements expressed in wavelengths and $f_{\mathrm{c}}$ is the center frequency.
To achieve beamforming, the modulating signal per antenna is cyclically shifted by $m d$ samples. The phase shift at the $m$-th antenna according to \eqref{eq:simpl_aliased_h_coef} is then 
\begin{align}
    \label{eq:phase_shift_ped_antenna}
    \phi_m(i,d) 
                &=-2\pi m \frac{d}{D} \left( 1 + \left( i + \left\lfloor \frac{A-1}{2} \right\rfloor \right)N \right).
\end{align}

The array factor (AF) of the TMA at the $i$-th harmonic frequency for a selected delay $d$ is 
\begin{align}
    \label{eq:tma_array_factor}
    \mathrm{AF}(\theta, i, d) = \alpha(i) \frac{1}{\sqrt{M}}  \sum_{m=0}^{M-1} & G_{m}(\theta) e^{j \phi_m(i, d)} \\
    & \times e^{-j 2\pi  \frac{\Delta d_m}{c} \left(f_{\mathrm{c}} + \frac{f_{\mathrm{p}}}{N} + i f_{\mathrm{p}} \right) } \nonumber .
\end{align}
where $G_{\mathrm{m}}(\theta)$ denotes the embedded/active element pattern~\cite{Balanis} and the factor $1/\sqrt{M}$ normalizes the total transmitted power to unity.
Assuming that the frequencies of the harmonic components with significant power are negligible compared to the carrier frequency $\left( \frac{1}{N} + i\right) f_{\mathrm{p}} \ll f_{\mathrm{c}}$ the AF can be simplified to
\begin{align}
    \label{eq:tma_af_simplified}
    \mathrm{AF}(\theta, i, d) = \alpha(i) \frac{1}{\sqrt{M}} \sum_{m=0}^{M-1} 
    &  G_{m}(\theta) e^{-2\pi m d \left( \frac{1}{D} + \frac{1}{O_{\tau}} \left( i + \left\lfloor \frac{A-1}{2} \right\rfloor \right) \right) } \nonumber \\
    & \times e^{-2\pi m d d_{\lambda} \sin{(\theta)} }.
\end{align}

Depending on the harmonic index and delay value, the formed beams are pointed in directions 
\begin{equation}
    \label{eq:beamforming_angles}
    \theta(i, d) = -\arcsin{ \left( \frac{d}{d_{\lambda}} \left(\frac{1}{D} + \frac{1}{O_{\tau}} \left( i + \left\lfloor \frac{A-1}{2} \right\rfloor \right) \right) \right)}.
\end{equation}

In the presence of pulse oversampling $O_{\tau} \geq 2$, the beamforming directions depend on the harmonic index $i$. When $d$ is not a multiple of $O_{\tau}$, the phase shift across the passband blocks $i \in i_{\mathrm{bb}_\mathrm{A}}$ is not the same, leading to different beamforming directions depending on the block index. As a result, only $ A\frac {d}{O_{\tau}}$ blocks are beamformed in the desired direction, reducing the effective bandwidth, while the remaining harmonics are beamformed in other directions. 
Fig. \ref{fig:ph_shift_per_d} shows the phase shift per harmonic index and the effect of varying phases across blocks when pulse oversampling is considered.

The pulse oversampling effectively improves the phase shifting resolution, and given the reduced switching frequency, it can be easily achieved. However, its feasibility is limited due to the loss of effective beamformed bandwidth and the requirement of an additional precoding vector and signal distribution for each pulse oversampling factor.

\subsection{System considerations}
\label{sec:sys_consideration}
When designing an ATMA system, the choice of the parameters $A$, $N$, $O_{\tau}$ is constrained. The phase shift of the block with index $a = i_{\mathrm{bb}} + \left\lfloor \frac{A-1}{2} \right\rfloor $ due to delay is given by 
\begin{align}
    \label{eq:ph_diff_per_d}
    \Delta \phi(a) &= \arg{(\alpha_{\mathrm{A}}(i_{\mathrm{bb}},d + 1))} - \arg{(\alpha_{\mathrm{A}}(i_{\mathrm{bb}},d))} \nonumber \\
    &= - 2 \pi \frac{1}{O_{\tau}} \left( \frac{1+aN}{N}\right)
\end{align}
To guarantee that each $a$ block (within $i_{\mathrm{bb}}$) observes the same relative phase shift, the following constraints arise
\begin{equation}
    \label{eq:multiple_of_constraint}
    1 + aN \neq p O_{\tau},\ p \in \mathbb{N}^+
\end{equation}
and
\begin{equation}
    \label{eq:gcd_constranint}
    \gcd \left( 1+aN, O_{\tau} \right) = 1.
\end{equation}
The $1+aN$ and $O_{\tau}$ must not have a common divisor for all considered $a$. The constraints are satisfied for all $a$ if both $O_{\tau}$ and $N$ are even; otherwise, limitations regarding $A$ occur.
Moreover, the number of samples corresponding to the modulation symbol duration must be a multiple of the modulating sequence length $N$. So that at least one complete cycle of time modulation occurs within the symbol duration. This requires upsampling of the modulation symbol at least by a factor $N$, which, for a single carrier system, further reduces the system's capacity. Due to the required extended duration of the modulation symbols in the time domain and the block structure of the transmitted signal, Orthogonal Frequency Division Multiplexing (OFDM) modulation is a natural choice for ATMA. 

\begin{figure}[b]
    \centering
    \includegraphics[width=\linewidth]{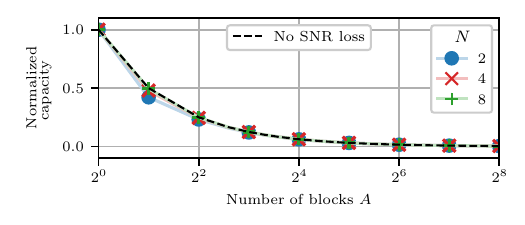}
    \caption{System communications capacity as a function of number of blocks $A$ for selected $N$ values.}
    \label{fig:atma_capacity}
\end{figure}

Consider an OFDM modulator with $K = A K_{\mathrm{b}}$ subcarriers, where $K_{\mathrm{b}}$ is the number of subcarriers per block. To satisfy the constraint that the integer number of time modulation cycles should happen within a single symbol duration, the number of subcarriers per block must be a multiple of the number of phase shift steps
\begin{equation}
    \label{eq:symb_dur_constraint}
    K_{\mathrm{b}} = p N,\ p \in \mathbb{N}^+
\end{equation}
The same applies to the length of a cyclic prefix if considered.
The general preference for choosing $K$ and $N$ as a power of 2 requires $O_{\tau}$ and $A$ also to be a power of 2.
The repeated block structure of the signal in the frequency domain reduces the communications capacity of the system as the aliasing factor $A$ increases. Due to aliasing, the symbol rate of the system of $B$ bandwidth is
\begin{align}
    \label{eq:symb_rate}
    R&=\frac{B}{A} \frac{K}{K + N_{\mathrm{cp}}},
\end{align}
where $N_{\mathrm{cp}}$ is the cyclic prefix length.
The repeated modulation symbols in the frequency domain offer improved outage probability, but combining them does not improve SNR due to increased noise bandwidth. The power fluctuations per block in the passband might result in different SNRs being received at each block, resulting in SNR and capacity loss. Note that repetition in the frequency domain increases the peak-to-average power ratio. In the presence of a nonlinear amplifier, the signal is more susceptible to distortion. Fig. \ref{fig:atma_capacity} presents the normalized capacity of the aliased TMA system. The values are normalized to a system without aliasing. The capacity loss due to SNR is negligible for $A$ and $N$ values greater than 4.

To successfully demodulate the ATMA OFDM, the receiver must revert the phase precoding per block applied at the transmitter \eqref{eq:alternating_precoder}.
The failure to do so would result in erroneous demodulation of $\left\lfloor A/2 \right\rfloor$ of the subcarrier blocks. 
Assume that the information regarding the precoding and the signal structure is available at the receiver or has been communicated prior. To revert the phase precoding, the receiver requires the information about the subcarrier block with $K_{\mathrm{b}}$. The repeated modulation symbols offer improved outage probability (e.g., robustness against deep fading). Combining $A$ repeated symbols offers an averaging gain and improved robustness against intercarrier interference due to uncompensated carrier frequency offset or phase noise.
However, the performance of synchronization algorithms will be negatively impacted by the repeated subcarrier structure. For example, when considering the Schmidl-Cox algorithm, $A$ times repetition of the subcarrier block will also shorten the correlation length $A$ times, resulting in a degraded estimate.

Another effect of aliasing that can impact the system performance is the per-block power variations. This is particularly important in the sensing application, where uncompensated power variations degrade the range profile. When considering OFDM sensing employing a zero-forcing estimator, the uncompensated amplitude variations will manifest as increased sidelobes in the range profile.
For communications, the demodulation and symbol detection would happen at the per subcarrier basis and the power variations would be absorbed in the conventional channel estimation and equalization.

The key parameters of an ATMA OFDM system are: symbol rate \eqref{eq:symb_rate}, switching frequency \eqref{eq:sw_freq}, phase shifting resolution \eqref{eq:num_phase_shifts} and ACLR \eqref{eq:aclr_def}. Consider some maximum supported switching frequency $f_{\mathrm{sw} \max}$ expressed as a fraction of bandwidth. The minimum aliasing factor is $A_{\min} = B / f_{\mathrm{sw} \max}$. Knowing the minimum aliasing factor, the maximum pulse oversampling factor can be calculated as follows $O_{\tau \max} = A / A_{\min}$. The maximum value of $A$ is not constrained, but increasing it reduces the symbol rate.
For example consider a system of bandwidth $B$, $N_{\mathrm{cp}} = 0.25K$, $N=4$ and and $f_{ \mathrm{sw} \max} = 1/4 B$. The $A$ and $O_{\tau}$ can be allocated in numerous ways. Tab. \ref{tab:example_sys_params} presents the system parameters for a few configurations of $A$ and $O_{\tau}$. Increasing $A$ enables higher values of the pulse oversampling at the cost of a significantly reduced communication rate.
\begin{table}[t]
    \caption{System parameters for different allocations of $A$ and $O_{\tau}$ for $N_{\mathrm{cp}} = 0.25K$, $N=4$ and $f_{ \mathrm{sw} \max} = 1/4 B$.}
    \label{tab:example_sys_params}
    \centering
    \def\arraystretch{1.2}
    \begin{tabular}{
      >{\centering\arraybackslash}m{0.25\linewidth}<{}
      |>{\centering\arraybackslash}m{0.1\linewidth}<{}
      |>{\centering\arraybackslash}m{0.1\linewidth}<{}
      |>{\centering\arraybackslash}m{0.05\linewidth}<{}
      |>{\centering\arraybackslash}m{0.2\linewidth}<{}
    }
         Allocation & $R\ [B]$ & $f_{\mathrm{sw}}\ [B]$ & $D$ & ACLR [dB] \\
         \hline
         $A=4$, $O_{\tau} = 1$ & $1/5$ & $1/4$ & $4$ & $19.71$ \\
         \hline
         $A=8$, $O_{\tau} = 2$ & $1/10$ & $1/4$ & $8$ & $22.04$\\
         \hline
         $A=16$, $O_{\tau} = 2$ & $1/20$ & $1/8$ & $8$ & $24.66$ \\
         \hline
         $A=64$, $O_{\tau} = 4$ & $1/80$ & $1/16$ &  $16$ & $30.34$ \\
    \end{tabular}
\end{table}
When considering the ATMA scheme in the context of the antenna array, each antenna is preceded by an identical time modulator front-end module. Therefore, any circuitry losses and imperfections will identically affect all antenna elements and scale linearly with the number of antennas.

The proposed system is suitable for communications-centric systems for low values of $A < 8$, which allows for retaining more than 10\% of the communication capacity offered by total system bandwidth $B$.
Based on Fig. \ref{fig:aclr_vs_phres_vs_nblocks}, satisfying the 5G ACLR requirement of 45 dB for a small aliasing factor ($A < 8$), requires high-resolution phase shifters ($N \approx 2^6$), increasing the hardware complexity. On the other hand, increasing the number of aliasing blocks $A$ relaxes the hardware requirement on $N$ at the cost of a significant reduction in the communication capacity, as shown in Fig. \ref{fig:atma_capacity}.
To maintain a reasonable communication rate while keeping the $N$ relatively small, some sort of front-end bandpass filtering needs to be considered. The band-pass filtering can be implemented via antenna design by limiting its frequency response or by adding a dedicated front-end filter.
For communication-centric applications, the proposed method should be viewed as a way to further relax the sideband filtering requirements.
For higher values of $A > 8$, it is more practical to consider the proposed method for radar-centric operation, where the communication capacity is of secondary importance. Due to the reduced communication rate, the system is envisioned to be used in radar-centric joint communications and sensing systems. 
Repeated data symbols lead to repetition peaks in correlation-based range estimation. In here, the zero-forcing (ZF) estimator is considered for range estimation. It removes the data dependence from the channel estimate and enables the use of full bandwidth for range processing.

\section{Experimental validation}
\label{sec:experimental_validation}

\subsection{Array design}
\label{sec:arr_design}

As a representative example, an eight-element linear TMA is designed with an inter-element spacing of $0.5\lambda_0$ at the reference frequency $f_0$~=~ 2.5~GHz. Each element is a double-dipole antenna, based on the design in~\cite{Double_Dipole_Antenna}, implemented with a partial ground plane. The antenna operates in the [2–3]~GHz band and is implemented on a 0.5-mm-thick Rogers~4350B substrate ($\epsilon_r = 3.66$, $\tan\delta = 0.0037$). It consists of two dipoles of different lengths, denoted as L$_1$ and L$_2$. The left halves of the dipoles are printed on the top substrate layer, while the right halves are placed on the bottom layer. These dipole sections are connected to a microstrip feed line with a truncated ground plane via printed microstrip lines on their respective layers. The partial ground plane serves as a reflector, producing an end-fire radiation pattern along the $y$-axis.

\begin{figure}
    \centering
    \includegraphics[width=0.6\linewidth]{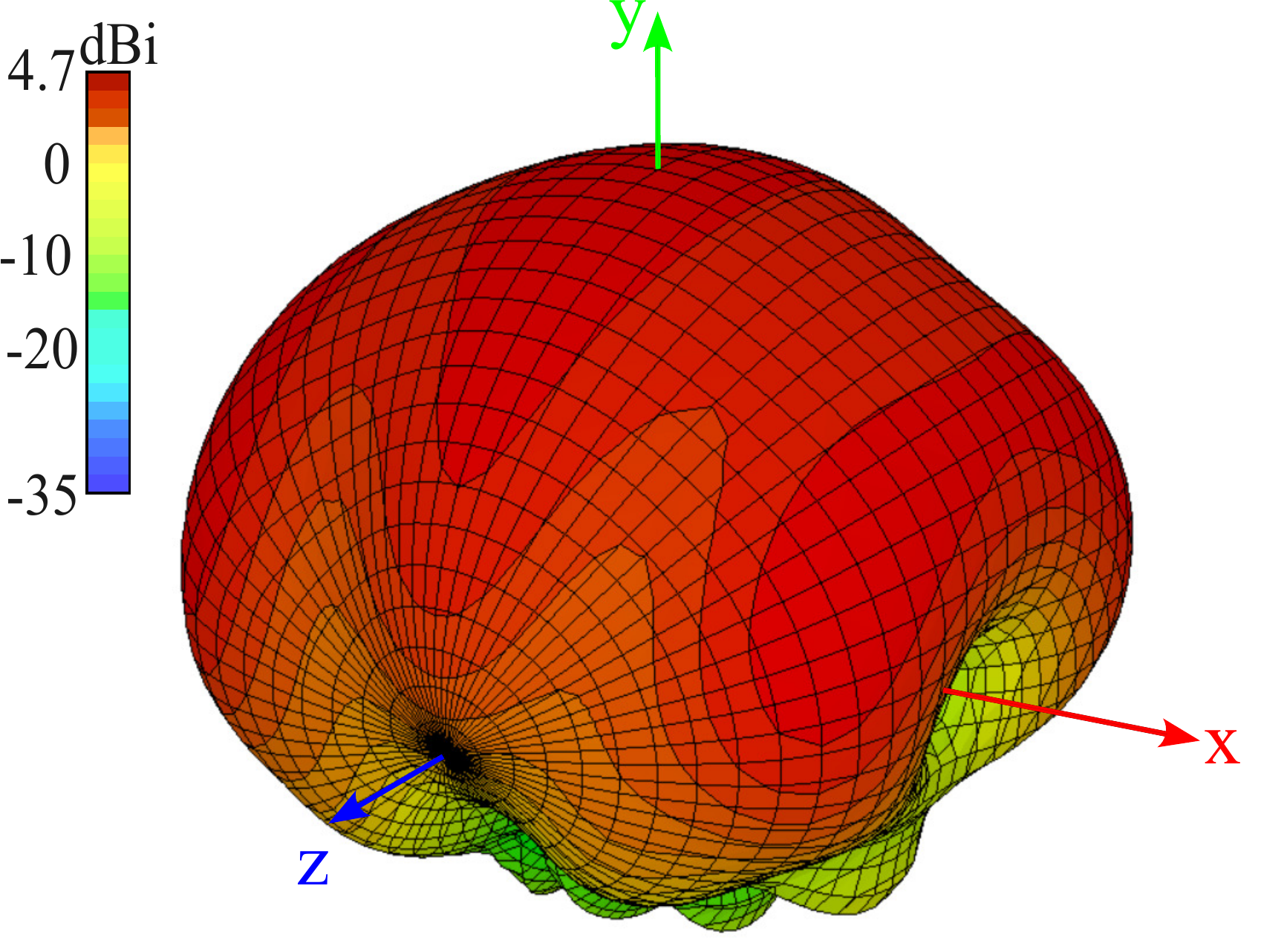}\\
    \small(a)\\
    \includegraphics[width=0.9\linewidth]{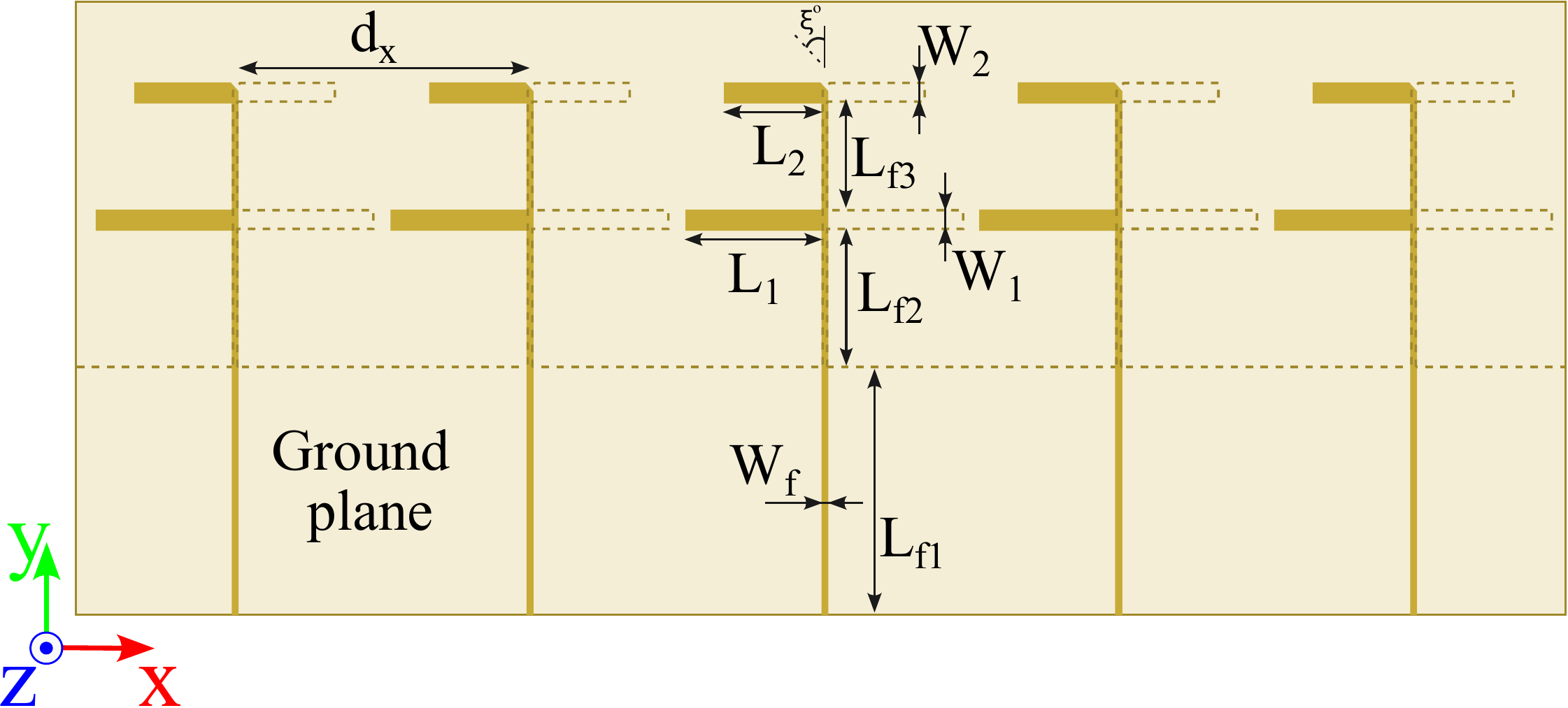}\\
    \small(b)\\ 
    \includegraphics[width=1\linewidth]{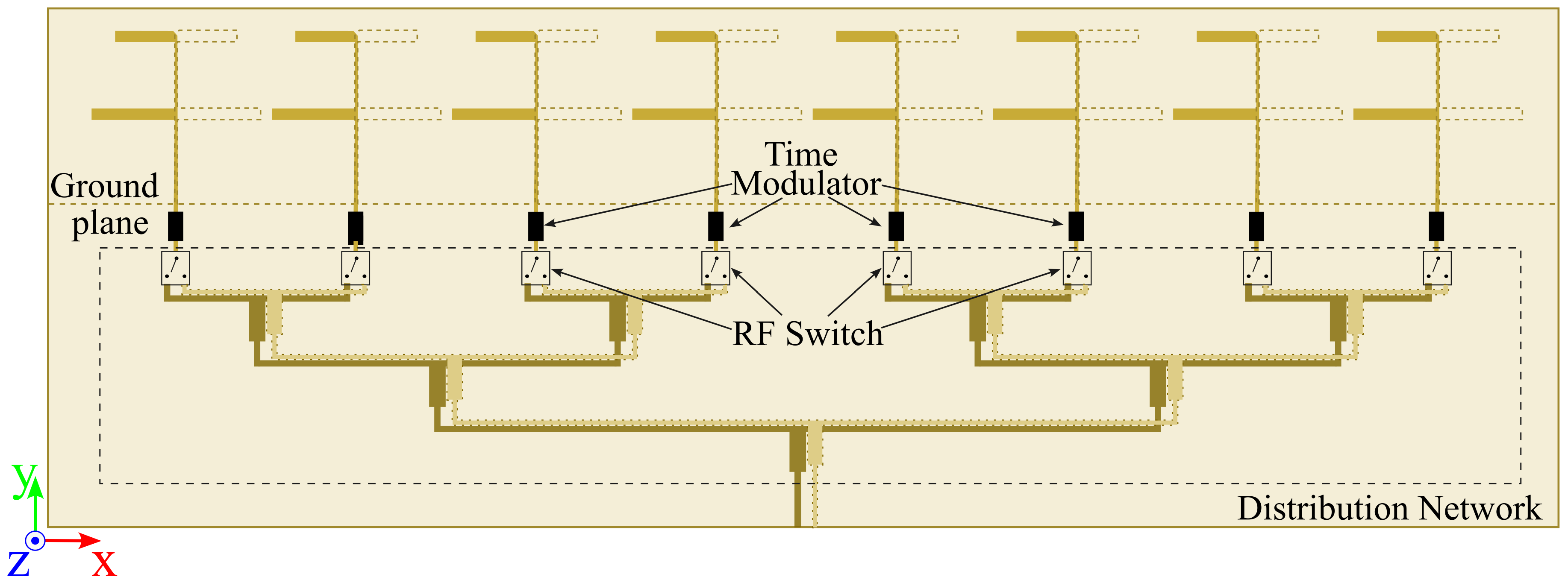}\\
    \small(c)\\    
    
    \caption{ Numerical Validation-3-D plot of (a)~embedded element pattern of the double dipole radiator at the central operation frequency of 2.5~GHz simulated in (b)~neighborhood of 5-element identical array elements. (c)~Sketch of the complete array model with the TMA feeding network, including time modulator and RF switches, simulated using a full-wave solver. Final dimensions: d$_\text{x}$~=~60~mm, L$_\text{1}$~=~28~mm,  L$_\text{2}$~=~20~mm, W$_\text{1}$~=~3~mm, W$_\text{2}$~=~3.75~mm, W$_\text{f}$~=~1~mm, L$_\text{f1}$~=~50~mm, L$_\text{f2}$~=~28.5~mm, L$_\text{f3}$~=~22.25~mm, $\xi$~=~45$^\circ$.}
    \label{fig:array_topology}
\end{figure}

To incorporate realistic radiating elements (i.e., $G_m(\theta) \neq 1$ $\forall m$), including the effects of mutual coupling, the radiation performance of the TMA is evaluated using full-wave simulations in CST Microwave Studio, accounting for interactions with two surrounding rings of neighboring elements [see Fig.~\ref{fig:array_topology}(b)]. The results remain consistent even when the neighbor set is further expanded, confirming that coupling effects are properly included, as illustrated in Fig.~\ref{fig:array_topology}(a)]. To assess the influence of the beamforming network (BFN), an equal-amplitude, equal-phase corporate-feed distribution network implemented in microstrip technology is integrated into the eight-element linear array along with time modulators and RF switches [see Fig.~\ref{fig:array_topology}(c)]. For conceptual clarity, a parallel distribution network, depicted in Fig.~\ref{fig:atma_diagram} and also included in Fig.~\ref{fig:array_topology}(c), is introduced. Yet, the full-wave solver simulations are carried out with only a single distribution network.

The EM simulations presented in this work are obtained using static excitations of the antenna elements, as described in Sec. \ref{sec:mod_design}. The full-wave simulation accurately models the radiation pattern of the array, including mutual coupling effects between antenna elements. The experimental measurements capture the time-domain behavior of the modulator and the resulting aliased OFDM signal, including the amplitude and phase of each subcarrier block. The measured signal is then used as the excitation for the EM simulations. Therefore, the EM simulations capture array-level radiation effects, while the measurements capture the signal distortions introduced by the time-modulation hardware. Effects such as dynamic switching behavior or RF hardware impairments are not directly modeled in the EM simulations but are implicitly included through the measured excitation signals.

\subsection{Modulator design}
\label{sec:mod_design}

\begin{figure}[t!]
    \centering
    \includegraphics[width=\linewidth]{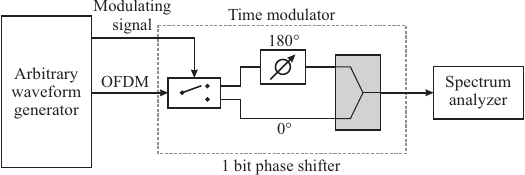}
    \caption{Diagram of the experimental setup.}
    \label{fig:atma_experiment}
\end{figure}

The feasibility of the proposed sideband reduction method has been experimentally validated with a one-bit phase shifter. For the experiment, a simple 1-bit phase shifter was constructed using a single-pole double-throw RF switch ZASWA-2-50DR+ from Mini-Circuits. The two switch outputs were connected to the combiner (ZFRSC-123+ from Mini-Circuits) directly and via a manual tunable phase shifter ARRA 9428A. The center frequency of the experimental setup was 2.5 GHz. The phase difference between the two switch-controlled paths was calibrated to 180$\degree$. The measured gain mismatch due to different path attenuation was around 0.1 dB and the phase shift difference error was below 1$\degree$. The precoded OFDM signal and the switching signal were generated by the Keysight M8190A arbitrary waveform generator. The signal at the output was captured with a Rohde \& Schwarz FSW26 signal and spectrum analyzer.
The diagram of the experimental setup can be found in Fig. \ref{fig:atma_experiment}. The proposed method is envisioned for multiple antenna systems; however, as each antenna is supposed to be equipped with identical time modulators, it is sufficient to demonstrate the performance of a single module.
The experimental measurements capture the time domain effects that are the core of generating an aliased OFDM signal. The measured output signal, including the phase and amplitude at each subcarrier block, then serves as an input for the EM simulation.

The measurement was performed for different numbers of blocks (aliasing factor) and two scenarios. In both scenarios, the center frequency of the transmitted passband signal was adjusted to compensate for the frequency offset due to time modulation. This results in spectra always centered at 2.5 GHz for ease of comparison. To guarantee high-resolution spectrum measurement, the number of subcarriers $K$ was 16384. For each measurement, the spectrum is normalized with regard to its maximum. The analytical results are also normalized to the maximum of each trace. In the first scenario, the transmitted signal bandwidth was constant while the switching frequency was adjusted to introduce aliasing for different numbers of blocks $f_{\mathrm{sw}} = B / A$. Note that the signal bandwidth is equal to the sampling rate $B=f_{\mathrm{s}}$ and the two parameters are used interchangeably. Fig. \ref{fig:meas_const_bw} compares the measured and theoretical values for the first scenario. As expected, increasing the number of blocks efficiently reduces the sideband radiation.

\begin{figure}[t!]
    \centering
    \includegraphics[width=\linewidth]{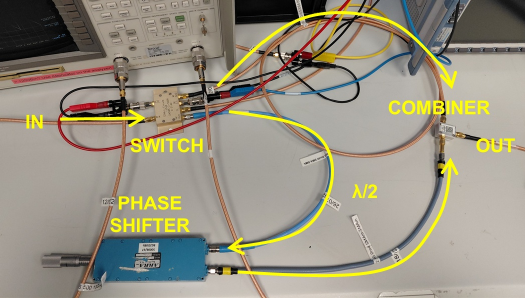}
    \caption{Picture of the experimental setup.}
    \label{fig:atma_experiment_lab}
\end{figure}

To validate the signal model and phase shifting improvement by pulse oversampling, the transmitted signal was demodulated. The transmitted signal consisted of a sequence of OFDM symbols to capture the phase shift introduced by the shifted modulating sequence. The measurements were performed for $O_{\tau} = 2$ and $4$. For each symbol, the precoder was adjusted based on \eqref{eq:prec_extended} and the modulating signal was cyclically shifted. Fig. \ref{fig:atma_ph_meas} presents the measurement of the phase shift per subcarrier block in the presence of pulse oversampling. As expected by \eqref{eq:simpl_aliased_h_coef} and also shown in Fig. \ref{fig:ph_shift_per_d}, the pulse oversampling effectively improves the phase shifting resolution. 
The measured EVM of the demodulated symbols generally increased with the growing number of blocks
$2\!:\!-28.2\,\text{dB}$,
$4\!:\!-31.8\,\text{dB}$,
$8\!:\!-30.7\,\text{dB}$,
$16\!:\!-29.0\,\text{dB}$,
$32\!:\!-25.0\,\text{dB}$,
$64\!:\!-24.3\,\text{dB}$,
$128\!:\!-21.9\,\text{dB}$,
$256\!:\!-20.6\,\text{dB}$.
With the increasing number of aliasing blocks, the power of the residual distortion caused by hardware imperfections accumulates, limiting the performance. Based on the EVM, the considered setup can support $A$ in the order of a few hundred before the performance degradation due to imperfect hardware and time modulation becomes too severe. 
Based on the measured EVM levels and assuming a high SNR regime where the system performance is limited by the EVM, the results indicate that for low $A<16$ the system can support 256-QAM, which requires EVM $<-29.1$ dB \cite{3gpp_ts_38_104_v1660}. However, with a larger number of blocks, the maximum supported order reduces to 64-QAM, which requires an EVM $< -21.9$ dB.

\begin{figure}[t]
    \centering
    \includegraphics[width=\linewidth]{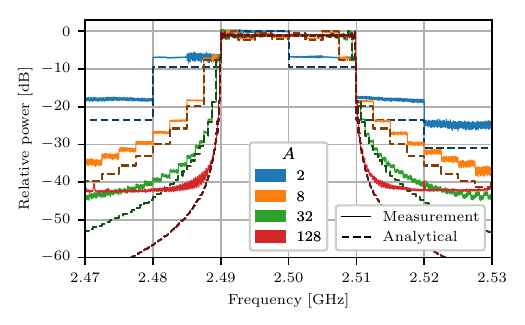}
    \caption{Measured spectrum of the ATMA for $N=2$, a constant signal bandwidth $B = 20$ MHz and different number of blocks. The switching frequency was adjusted as $f_{\mathrm{sw}} = B/A$.}
    \label{fig:meas_const_bw}
\end{figure}
\begin{figure}[t!]
    \centering
    \includegraphics[width=\linewidth]{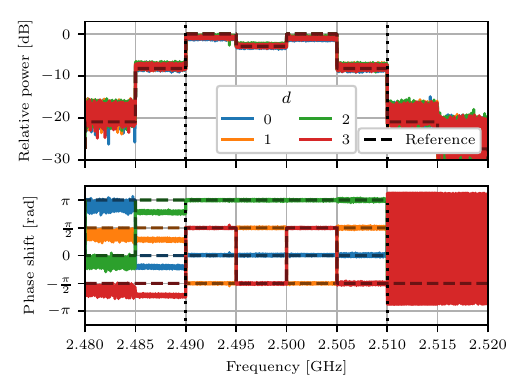}
    \caption{Measured spectrum and phase shift per block for $N=2$, $A=4$ blocks, pulse oversampling $O_{\tau} = 2$ and signal bandwidth $B = 20$ MHz. The switching frequency was $f_{\mathrm{sw}} = B/A$. The phase shift was measured with regard to $d=0$.}
    \label{fig:atma_ph_meas}
\end{figure}

\begin{figure}[t]
    \centering
    \includegraphics[width=\linewidth]{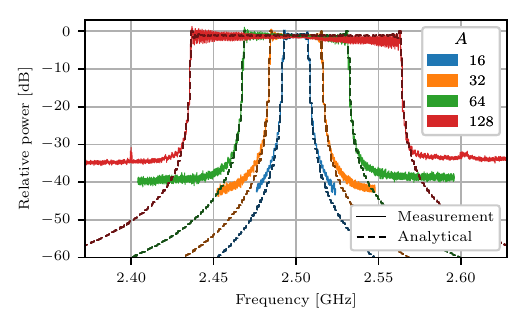}
    \caption{Measured spectrum of the ATMA for $N=2$, a constant switching frequency $f_{\mathrm{sw}} = 1$ MHz and different number of blocks. The signal bandwidth was adjusted as $B = A f_{\mathrm{sw}}$.}
    \label{fig:meas_const_fsw}
\end{figure}

In the second scenario, the switching frequency was constant $f_{\mathrm{sw}} = 1$ MHz and the signal bandwidth was adjusted with the number of blocks to introduce aliasing $B = A f_{\mathrm{sw}}$. This scenario is devised to present the wideband performance of the proposed method. As shown in Fig. \ref{fig:meas_const_fsw}, the method supports wideband signals. 
The wideband signals are supported in the following sense: the original transmitted signal is wideband with bandwidth $B$. To enable switching frequencies lower than the signal bandwidth and retain the phase shifting offered by time modulation, the original signal is designed to be robust against aliasing. This is achieved by introducing repeated subcarrier blocks and precoding that facilitates constructive combining of the harmonics within the original signal bandwidth $-B/2$ to $B/2$, effectively reconstructing the original baseband signal after the time modulation.

Based on the measurements, the maximum bandwidth supported by the experimental setup is in the order of a few hundred MHz, for example, $A = 128$ and $f_{\mathrm{sw}} = 1$ MHz. The further increase in the number of blocks does not yield expected gains in the sideband attenuation due to the accumulation of the distortion due to hardware imperfections.

A discrepancy between the measured and analytical power levels per block can be observed for both scenarios. This can be attributed to the imperfections of the hardware, such as phase and amplitude imbalance between the phase shifter states, leading to imperfect aliasing and resulting in the presence of residual power. The amplitude imbalance between the states results in an asymmetrical waveform and introduces a DC component in the harmonics generated by the switching. The limited bandwidth supported by the switch and the non-negligible switching transients introduce tapering of the harmonic powers as their frequency increases. All those effects add up to the total performance degradation and imperfect aliasing of the blocks. Mentioned imperfections scale with the switching frequency and the number of blocks, posing a practical constraint on the number of aliasing blocks.

\subsection{Results and Comparison}

\begin{figure}[t]
    \centering
    \includegraphics[width=\linewidth]{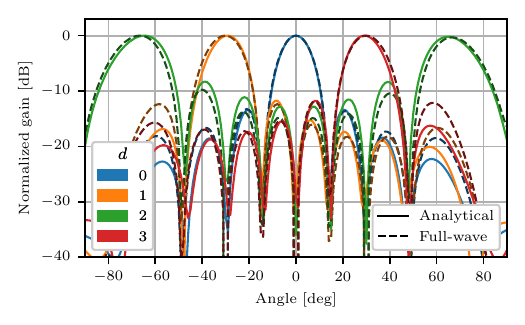}
    \caption{Comparison of the EM simulated and analytical beampattern per delay for the first $i = -1$ subcarrier block for $N=2$, $A=4$ blocks, pulse oversampling $O_{\tau} = 2$ and total signal bandwidth $B = 20$ MHz.}
    \label{fig:em_beampattern_comp}
\end{figure}

To validate the beamforming capability of the ATMA the results obtained from the measurement of the modulator from Sec. \ref{sec:mod_design} were incorporated into the full-wave simulation of the array designed in Sec. \ref{sec:arr_design}.
To provide a representative example the number of blocks is $A=4$ with total signal bandwidth $B = 20$ MHz and pulse oversampling $O_{\tau} = 2$. The remaining parameters are the same as in the preceding sections \ref{sec:mod_design}, \ref{sec:arr_design}.
Fig. \ref{fig:em_beampattern_comp} presents the comparison between the full-wave EM simulated beampattern and the analytical formulation from \eqref{eq:tma_array_factor}. As can be seen, the analytical model is in good agreement with the full-wave simulations, with minor discrepancies in the sidelobes and beampointing direction.

The Table \ref{tab:sota_comparison} presents the performance of the proposed method in comparison to the state-of-the-art. For the experimental setup, the maximum measured SR was $-$6.57 dB for $N=2$ and $A=128$. 
Due to the stair-like roll-off of the SR in the ATMA the maximum SR is determined by the first block adjacent to the desired signal. The power of the first adjacent blocks is mainly determined by the number of phase shifts $N$, see Fig. \ref{fig:atma_spectrum}. As the number of aliasing blocks $A$ increases, the width of the block is reduced. However, the power of the nearest blocks remains relatively high, due to an imbalance in the resulting phase of the aliased (summed) blocks. As can be seen in Fig. \ref{fig:meas_const_bw}, the nearest SR blocks introduce high maximum SR, even though their width is almost negligible. 

The aliasing improves the SR over a wide bandwidth by averaging oppositely precoded blocks.
The maximum SR of ATMA is greatly improved when some transition bandwidth is considered. The transition bandwidth allows for taking the SR roll-off into account. For the transition bandwidth of 1\%, the maximum measured SR is $-$18.66 dB and for 10\% it becomes $-$36.32 dB. When considering transition bandwidth, the SR performance of the proposed work is comparable to the state-of-the-art, with the advantage that it requires much lower switching frequency and simple hardware. The binary amplitude modulations \cite{sb_supp_nonuniform, sb_suppresion_tma_freq, tma_cont_spectrum} offer great SR suppression at the cost of poor efficiency and high switching frequency. Similarly, the IQ modulation \cite{sb_supp_stair_pulses, sb_supp_for_wide_bandwidth, iq_ssb_tma, iq_modulation_enhanced_tma} offers good sideband attenuation at the cost of complex hardware requiring high-resolution phase and amplitude control.
The proposed aliasing method matches the performance of the state-of-the-art solutions when transition bandwidth is considered, ultimately trading off the spectral efficiency for sideband attenuation.

\section{Conclusion}
\label{sec:conclusion}
The proposed ATMA architecture enables the reduction of sideband radiation and switching frequency by splitting the baseband signal into repeated blocks, which are differently precoded and subsequently aliased. The number of blocks is an additional dimension that can be tuned along with the phase shifter resolution $N$ to reduce the sideband radiation and allow the TMA to meet the ACLR criteria of a conventional system while having a simpler architecture. The pulse oversampling with aliased TMA requires an additional precoder for each pulse oversampling factor $O_{\tau}$, limiting the practicality of large pulse oversampling. Since the scheme preserves the total bandwidth but reduces the spectral efficiency, it is well-suited for radar applications as well as radar-centric integrated communication and sensing systems.
The feasibility of ATMA was validated by experimental measurement and full-wave simulations.

\bibliographystyle{IEEEtran}
\bibliography{IEEEabrv.bib, biblio.bib}

\end{document}